\documentclass[sigconf]{acmart}
\usepackage[utf8]{inputenc}
\usepackage[mathscr]{euscript}
\usepackage{paralist}
\usepackage{booktabs}
\usepackage{hyperref}
\usepackage{cleveref}
\usepackage{color}
\usepackage{soul}
\usepackage{xspace}
\newcommand\etal{et\,al.\xspace}
\newcommand\ie{i.\,e.\xspace}
\newcommand\eg{e.\,g.\xspace}
\usepackage{amssymb}
\usepackage{graphviz}

\definecolor{amber}{rgb}{1.0, 0.49, 0.0}
\definecolor{INFO}{RGB}{224,201,173}

\newcommand{\paperortr}[2]{#1} 

\begin{document}
\setcopyright{acmcopyright}




%

\acmConference[K-CAP 2017]{Knowledge Capture}{December 4th-6th, 2017}{Austin, Texas, United States}
 
\paperortr{%
  \title{Using Titles vs. Full-text as Source for\\ Automated Semantic Document Annotation}
}{%
\title{Comparing Titles vs. Full-Text for Multi-Label Classification of Scientific Papers and News Articles}
}

\renewcommand{\shorttitle}{Titles vs. Full-text for Automated Semantic Annotation}

\author{Lukas Galke}
\affiliation{%
  \institution{ZBW---Leibniz Information Centre for Economics, Kiel}
}
\orcid{0000-0001-6124-1092}
\email{l.galke@zbw.eu}

\author{Florian Mai}
\affiliation{%
  \institution{Kiel University}
}
\email{stu96542@mail.uni-kiel.de} 

\author{Alan Schelten}
\affiliation{%
  \institution{Kiel University}
}
\email{stu111405@informatik.uni-kiel.de}

\author{Dennis Brunsch}
\affiliation{%
  \institution{Kiel University}
}
\email{deb@informatik.uni-kiel.de}

\author{Ansgar Scherp}
\affiliation{%
  \institution{ZBW---Leibniz Information Centre for Economics, Kiel}
}
\email{a.scherp@zbw.eu}

\renewcommand{\shortauthors}{L. Galke, F. Mai, A. Schelten, D. Brunsch, A. Scherp}

\date{\today}


\begin{abstract} 
A significant part of the largest Knowledge Graph today, the Linked Open Data cloud, consists of metadata about documents such as publications, news reports, and other media articles. While the widespread access to the document metadata is a tremendous advancement, it is yet not so easy to assign semantic annotations and organize the documents along semantic concepts. Providing semantic annotations like concepts in SKOS thesauri is a classical research topic, but typically it is conducted on the full-text of the documents. For the first time, we offer a systematic comparison of classification approaches to investigate how far semantic annotations can be conducted using just the metadata of the documents such as titles published as labels on the Linked Open Data cloud. We compare the classifications obtained from analyzing the documents' titles with semantic annotations obtained from analyzing the full-text. Apart from the prominent text classification baselines kNN and SVM, we also compare recent techniques of Learning to Rank and neural networks and revisit the traditional methods logistic regression, Rocchio, and Naive Bayes. The results show that across three of our four datasets, the performance of the classifications using only titles reaches over 90\% of the quality compared to the classification performance when using the full-text. Thus, conducting document classification by just using the titles is a reasonable approach for automated semantic annotation and opens up new possibilities for enriching Knowledge Graphs.
\end{abstract}
 


%
%

%



\maketitle

\section{Introduction}\label{sec:introduction}

A significant amount of today's largest Knowledge Graph on the web, the
so-called Linked Open Data cloud\footnote{See latest version from 02/2017:
\url{http://lod-cloud.net/versions/2017-02-20/lod.svg}}, consists of metadata
about documents such as scientific papers and news articles. Domain-specific
SKOS vocabularies are used to describe the semantics of these documents, SKOS
(short for: Simple Knowledge Organization System)\footnote{SKOS:
\url{https://www.w3.org/2004/02/skos/}} is an established W3C standard for
modeling thesauri in domains such as economics, politics, social sciences,
news, etc. Those thesauri are often of high quality since they are manually
crafted as well as maintained by domain experts, and made freely available on the
web\footnote{An overview of current SKOS vocabularies is maintained by the
W3C: \url{https://www.w3.org/2001/sw/wiki/SKOS/Datasets}}.

The challenge is to successfully use those SKOS thesauri to semantically annotate the documents. 
However, the full-text PDF of the documents may not be available (linked from the documents' metadata) or may not be legally accessible due to licensing or copyright issues (even though there is a link to the PDF).
Thus, it is highly desirable to conduct a semantic annotation of the documents with the SKOS thesauri by just using the already published documents' metadata like the title, year, authors, etc.
In contrast to the full-text of documents, the metadata is directly available on the Linked Open Data cloud, accessible in RDF format, and can be processed with no legal barriers for semantic annotation. 
Conducting semantic annotations by using only the title (or further metadata of the documents) is challenging, since the title is short and thus carries only little information compared to the full-text.
The process of semantic annotation is a multi-label classification task where
not only one label is to be chosen as annotation but a set of labels since
many concepts of the SKOS thesauri are needed to appropriately describe the
semantics of the documents.

We tackle the challenge of conducting a semantic multi-label classification into SKOS thesauri by using only the title metadata of the documents.
To this end, we run an extensive series of experiments to compare established methods and recent methods from machine learning for document classification.
The goal is to decide whether it is possible to reach a comparable classification performance when using only the title of the documents.
It is noteworthy that all the compared approaches operate on the underlying machine
learning level which makes a comparison with prevalent end-to-end
ontology tagging systems such as SOLR ontology
tagger\footnote{\url{https://www.opensemanticsearch.org/solr-ontology-tagger}}
and MAUI\footnote{\url{https://github.com/zelandiya/maui-standalone}} difficult.
We instead show that despite not using the
hierarchical properties of the thesaurus, the presented methods
outperform the best-performing methods that do make use of the hierarchy such as
the ones of our own prior work~\cite{grosse2015comparison}.
Apart from the well-known multi-label classification baseline $k$-nearest
neighbors (kNN) and support vector machines (SVM), we revisit traditional text classification methods such as Naive
Bayes, Rocchio, and logistic regression (LR).
We also include the prominent Learning to Rank (L2R) approach, as well as a
modern variant of neural networks motivated by the success of the Deep
Learning field.
Please note, the present work focuses solely on using the titles of documents,
since they are the richest metadata attribute and contain keywords relevant in
the domain. In the future, we may also incorporate other metadata like authors'
names and publication year.
 
The results of our experiments show that it is possible to reach a competitive
performance for semantic annotation using solely the title of documents,
compared to exploiting the full-text of the documents.
Using a sample-averaged $F_1$ measure as evaluation metric, we compare the
automated predictions of semantic annotations from different methods with those
annotations provided by domain experts. We run our experiments over four
large-scale documents corpora of different origin and domain with a total of
over $300,000$ documents.
All datasets offer professional labels, i.\,e., manual annotations from domain
experts.
Two datasets are from professional scientific libraries in economics and politics while the other two datasets are the well-known news corpora from New York Times and Reuters.
In the past, algorithms of the lazy learner family such as kNN used to dominate multi-label classification tasks on such datasets with a high amount of classes~\cite{spyromitros2008empirical, grosse2015comparison}.
However, we show that eager learners such as logistic regression and feed-forward neural networks outperform lazy learners. 
Most eager learners have the benefit of $\mathcal O (N_{\text{parameters}})$ time complexity to predict a label
set for an unseen document, which is important when applying an automated
semantic annotation process for on-the-fly enrichment of metadata on the Linked
Open Data cloud. In contrast, lazy learners as well as Learning to Rank need to
store and traverse $\mathcal O(N_{\text{training examples}} \cdot
N_{\text{features}})$ space to predict the labels for a single new document at
test time. 
%
Finally, focusing on the
metadata also allows direct processing of data in published RDF format (\eg{}
the \textsf{rdfs:Literal} and \textsf{rdfs:label} information) without accessing the full-text of the documents at all. Overall,
we conclude that eager learning algorithms are well-suited for automated
semantic annotation of RDF resources in Linked Data.
Summarized, the contributions of this work are:
\begin{enumerate}
  \item To the best of our knowledge, the first large-scale systematic comparison of multi-label classifiers applied to either the full-text or only the titles of documents.
  \item Results that show that eager learners such as neural networks and linear models outperform lazy learners even when a high amount of possible labels is considered.
  \item We offer evidence that using only the title for high-dimensional multi-label classification is a reasonable choice for semantic annotation of resources where only metadata is available, such as documents modeled in RDF on the Linked Open Data cloud.
\end{enumerate}
%
%

The remainder of the paper is organized as follows: Below, we present an
overview of the state of the art in multi-label classification of text and
related fields. In \Cref{sec:overview}, we describe our experimental
apparatus. We depict different methods for conversion of unstructured text to
feature vectors in \Cref{sub:vect}. The classifiers and their respective
configurations are elaborated in detail in \Cref{sub:clf}. We describe the
four datasets used for our experiments as well as the evaluation metrics in
\Cref{sec:experiments}. The results are presented in \Cref{sec:results} and
discussed in \Cref{sec:discussion}, before we conclude. 

\section{Related Work}\label{related_work}

Most earlier work on the multi-label classification task with many possible
output labels relies on nearest neighbor searches (kNN). Using the union of
labels as well as separately voting for each individual label among neighbors
is a common choice in these nearest neighbor-based
classifiers~\cite{zhang2007ml,spyromitros2008empirical,DBLP:journals/jdwm/TsoumakasK07,DBLP:reference/dmkdh/TsoumakasKV10,
grosse2015comparison}.
Concept extraction~\cite{goossen2011news} refers to explicitly finding known
concept-specific phrases in the documents.
The extracted concepts are re-weighted by inverse document frequency, as in the
well-known TF-IDF~\cite{salton1988term} retrieval model. In our prior
work~\cite{grosse2015comparison}, we have conducted an exhaustive comparison of
concept extraction and feature re-weighting methods using kNN as a multi-label
classifier. 

Recent progress in the field of topic modeling with latent Dirichlet
allocation~\cite{blei2003latent} suggest using labeled
variants~\cite{DBLP:conf/emnlp/RamageHNM09, DBLP:conf/ic3k/BaiW15,
Soleimani:2016:SMT:2983323.2983752} for multi-label classification. While these
techniques outperform SVMs, we found from pre-experiments that they do not
scale well regarding the number of considered labels. 
In the closely related field of (label) recommendation, Tuarob
\etal~\cite{tuarob2013automatic} as well applied topic models to obtain a
ranking of the labels.

In the biomedical domain, the most popular approach is Learning to Rank~\cite{huang2011recommending,DBLP:journals/bioinformatics/PengYWZMZ16}.
The algorithm learns a ranking of the MeSH terms.
In multi-label classification, however, a hard decision is necessary to enable fully automated classification.
Thus, Learning to Rank is typically adjusted for multi-labeling by imposing a hard cut-off.
There are also approaches that use Learning to Rank along with dynamic cut-off techniques~\cite{DBLP:journals/biomedsem/MaoL17}.
The most prominent approach to adapt classifiers for multi-labeling is binary
relevance~\cite{TANAKA201585,DBLP:journals/jdwm/TsoumakasK07}. Other options include the
chaining~\cite{DBLP:journals/ml/ReadPHF11} as well as
stacking~\cite{hess2008multi, DBLP:conf/www/TangRN09} of classifiers. While the
former is not
well-suited for high amounts of considered labels, we also include a variation of
the latter idea in our comparison.
Bi and Kwok~\cite{DBLP:conf/icml/BiK13} approach the multi-label classification
task from a different direction. They strive for more efficient multi-label
classification and proper treatment of label correlation by transforming the
label indicator matrix.


Zhang and Zhou~\cite{zhang2006multilabel} have proposed to train a separate
neural network for each label along with a dedicated loss function. However,
this approach does not scale to high amounts of possible output labels. One
year later, the same authors suggest a lazy-learning multi-label variant of
kNN~\cite{zhang2007ml}, which is considered in our comparison. Nam
\etal{}~\cite{nam2014} adapt fully connected feed-forward neural networks for
multi-label classification by learning a threshold that determines whether a
label should be assigned or not.

While the related fields of label recommendation and single-label text
classification are broad, only few works consider multi-label classification
with a large amount of possible output labels. From these, the dominant
approaches are based on nearest neighbors searches, \ie{} lazy learners and
Learning to Rank. The considered works all use either short texts or full-text
as input data but do not compare these two different input variants. Thus, we
offer the first systematic comparison of text vectorization methods and lazy as
well as eager learning algorithms for the multi-label classification problem
with many possible labels applied to either title data or full-text data.

\section{Semantic Annotation Apparatus}\label{sec:overview}
\begin{figure}[t!]
  \centering 
  \includegraphics[width=1\columnwidth]{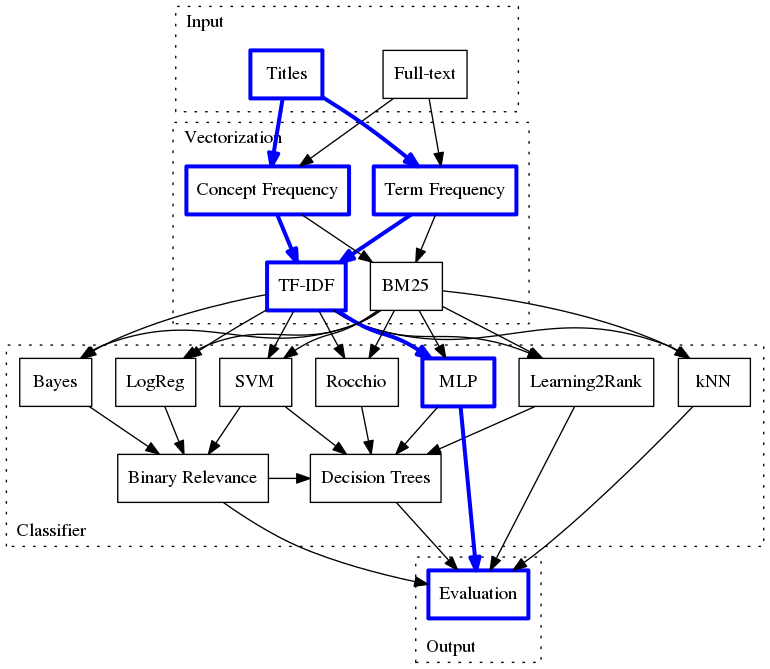}
  \caption{Illustration of the configurable text-processing pipeline used for
our experiments. The pipeline starts with the vectorization of the input text,
followed by feature re-weighting, classification and evaluation. The emphasized
edges and nodes show the most successful strategy applied to title
data.}\label{fig:pipeline} \end{figure}

We present an end-to-end apparatus for semantic annotation of unstructured
text. Figure~\ref{fig:pipeline} shows our generic text processing pipeline that
we used for the experiments. Each path through the graph resembles a possible
configuration.
%
%
In the following \Cref{sub:vect}, we describe the conversion from
unstructured text to a vector representation. In \Cref{sub:clf}, we elaborate
in detail on the classification methods that we have compared.
\subsection{Vectorization}\label{sub:vect}

  
%

\paragraph{Counting terms and extracting concepts}

In the first step of our text processing pipeline, the raw text needs to be converted into a
vector representation that can be supplied as input to the classifiers. As
features, we use the counts of \emph{term} occurrences in the text (TF) as well as
the number of times a \emph{concept} provided by a domain specific thesaurus
can be extracted from the text (CF). A concept is a set of concept-specific
phrases. In case of SKOS format, each concept has one preferred phrase
(\texttt{skos:prefLabel}) and optionally a set of alternative phrases
(\texttt{skos:altLabel}). We extract these concept-specific phrases from the
text using a finite state machine. When there is more than one possible match
in a sequence of words, we favor the longest phrase. We assume that longer
phrases carry more specificity. Hence, the occurrences of a concept (set of concept-specific
phrases) are counted in the same way as term occurrences.
The effect of concept extraction is to ensure that domain-specific synonyms
encoded in the thesauri are mapped to the same concept. The concepts are also
directly associated to the respective class labels. 
Hence it is left to the learning algorithm, to decide about the concrete label
assignment, given the extracted terms or concepts.

%

\paragraph{Discounting frequent terms and concepts}\label{subsec:IDF}
Inverse document frequency (IDF) is a re-weighting scheme introduced in the
1970s by Salton and Buckley~\cite{salton1988term} which has proven to work well for
information retrieval~\cite{manning2008introduction}. IDF discounts features that
occur in many documents of the corpus, and thus do not hold discriminative information.
This can be both term counts and counts of extracted concepts.
Let $D$ be the set of documents, then the IDF re-weighted score for some term
or concept $w$ in a document $d \in D$ is defined as:
$\text{TF-IDF}(w, d) = \text{TF}(w, d) \cdot \text{IDF}(w, D)$,
where $\text{IDF}(t,D)=1 + \log{\frac{\left\vert D \right\vert + 1}{\left\vert \left\{ d \in D : w \in d \right\} \right\vert + 1}}$.
To avoid division by zero, both the nominator and the denominator are incremented by one,
as if there was one artificial document containing all possible terms and concepts.
This can happen because the set of concepts given by the thesaurus but the data
itself might not cover all of these possible concepts.
The fraction as a whole is as well incremented by one, to ensure that words
that appear in all documents are not completely discarded.

Okapi BM25 is an extension of IDF by Robertson \etal~\cite{robertson1999okapi}
that slightly modifies the IDF term to include the average length of a
document. It offers two hyper-parameters for interpolating the difference
between the current document length and the corpus-wide mean document length.
The literature suggests to use BM25 especially for fields with short texts
using hyper-parameters $k=1.6$ and $b=0.75$~\cite{manning2008introduction}.
Hence, variants of our text vectorization methods using BM25 instead of TF-IDF
re-weighting are included in our comparison. 

\paragraph{Combining terms and concepts}
After re-weighting by either inverse document frequency or BM25, the resulting
vectors are normalized to unit length (with respect to the L2-norm). This leads to desirable invariance to
document length. Besides using only either the term frequency (TF) or the concept frequency (CF), we
concatenate the respective feature vectors (CTF).

\subsection{Classification}\label{sub:clf}
In the second step of the pipeline, a classifier is consulted to predict the
desired set of labels based on the vector representation of the input text
(compare \Cref{fig:pipeline}). Given training data, the classifiers have the
opportunity to learn how to associate the features with the respective class
labels. Lazy learners merely copy their input at training time, shifting the
main computational effort to test time (described in \Cref{sub:lazy}).  On the
other hand, eager learners use the training data for adapting their parameters
according to the correct classification result.  We describe those in detail in
\Cref{sub:eager}.  Some of the learning algorithms are only designed for
single-label classification (SVM, logistic regression, Naive Bayes), others do
only return a ranked list of possible labels (kNN, Rocchio, Learning to Rank).
We describe the multi-label adaption strategies for both cases in \Cref{sub:multi}.

%


\subsubsection{Lazy Learners}\label{sub:lazy}

\paragraph{Nearest Neighbor Classifier} The most typical lazy-learning
algorithm is $k$-nearest neighbors (kNN).  All training examples are stored
along with their class annotations.  At test time, the $k$ nearest neighbors
with respect to some distance metric (we chose cosine) vote on class
membership.  For multi-label problems, variants are proposed that assign the
union of label annotations in the neighborhood as well as conducting a separate
vote for each label~\cite{spyromitros2008empirical}.  By auto-optimizing the
$k$ hyperparameter for these methods, we found $k=1$ to be the optimal value in
our setting (as in our prior work~\cite{grosse2015comparison}). In this case
all multi-label variants coincide to copy the label set from the nearest
neighbor of the training set.


\paragraph{Rocchio Classifier}
The Rocchio classifier, or nearest-centroid classifier resembles a light-weight modification of the nearest neighbor classifier.
During training, only the centroid of each class is stored.
The classification result is then determined by the nearest of these centroids at test time.
In multi-label classification however, the classifier is only capable to return
a ranked list of labels based on the distance to the respective centroids.
As in the nearest neighbor classifier above, we use cosine distance as criterion.

\subsubsection{Eager Learners}\label{sub:eager}

\paragraph{Naive Bayes} The Naive Bayes classifier is one of the most
traditional classifiers for text classification tasks. We consider two Naive
Bayes variants, multinomial and Bernoulli. In the multinomial variant, the
features of term or concept frequencies are assumed to be generated by a
multinomial distribution. The Bernoulli variant only takes the occurrences of
(binary) features into account, which leads to penalizing the non-occurrences
of features. The Bernoulli variant is an intuitive approach for short text such
as titles since duplicate words are rather infrequent, while the multinomial
variant is more intuitive for full-texts. For both variants, we apply
Lidstone-Smoothing with $\alpha = 10^{-5}$. The main drawback of Naive Bayes
is the assumption of statistical independence among the input features.
 
\paragraph{Linear Models}\label{sec:GLM}
Generalized linear models~\cite{10.2307/2344614} use the training examples to
learn a decision boundary. This decision boundary is a separating hyperplane specified by a linear
combination of the input features $\mathbf{w} \cdot \mathbf{x} - b = 0$.
The parameters $\mathbf{w}$ and $b$ are optimized to minimize the regularized training error:
$\frac1n \sum_{i=1}^n J(y_i, y(\mathbf{x_i})) + \alpha R(\mathbf{w})$ 
where $y(\mathbf{x}) = \mathbf{w} \cdot \mathbf{x} - b$ is the model's output
and $\alpha R(\mathbf{w})$ is a regularization term on the model's weights
such as the L2-norm. 
For the loss function $J$, we consider two variants: logistic loss
$J_{\text{logistic}}(y,p) = \ln (1 + \exp(-p \cdot y))$
 as in logistic regression (\emph{LR}) and hinge loss $J_{\text{hinge}}(y,p) =  \max (0, 1 - p \cdot y)$
as in linear support vector machines (\emph{SVM}).
At test time, the binary decision is determined by the side of the hyperplane,
on which the document in question falls.
We employ stochastic gradient descent as an optimizer for
these generalized linear models, which is known to yield good generalization on
large-scale
datasets~\cite{zhang2004solving,bottou-bousquet-2008,bottou2010large}.
%
%
We apply the learning rate schedule $\eta^{(t)} = \frac1{\alpha \cdot (t_0 +
t)}$, where $t_0$ is chosen by a heuristic of Léon
Bottou~\cite{DBLP:series/lncs/Bottou12}. We average the weights $\mathbf{w}$
over time, which allows higher learning rates and leads to faster
convergence~\cite{DBLP:series/lncs/Bottou12}.
In this setting, we empirically determined $\alpha=10^{-7}$ to be a good
hyper-parameter value for all datasets (in the range $10^{-1}, 10^{-2}, \dots,
10^{-9}$). This leads to comparatively high initial learning rates and low
regularization.

\paragraph{Learning to Rank}\label{sec:l2r}
Learning to Rank (L2R) refers to a set of techniques that can be used to learn the
ranking of a list from training data.
As suggested by Huang \etal~\cite{huang2011recommending}, we restrict the supplied list to those
labels that occur in the $k$ neighboring documents (we empirically determined $k=45$). 
Those labels, that are also assigned to current document in question should be ranked higher than the others.
To learn the ranking, we use the neighborhood, overlap, and translation-probability features proposed by Huang~\etal~\cite{huang2011recommending}.
Hence at test time, the union of labels among the $k$ nearest neighbors are ranked via the learned parameters.
However, the algorithm itself does not offer
the possibility of hard decisions on label assignments. Thus, we chose to
cut off the ranked list at the position of the average number of assigned
labels in the training data. In our experiments, we made use of the
RankLib library\footnote{\url{https://people.cs.umass.edu/~vdang/ranklib.html}}
and found LambdaMART to outperform other list-wise L2R algorithms.

\paragraph{Multi-Layer Perceptron}\label{sec:MLP}  As representative for the
neural network family, we employ a fully connected feed-forward neural network
with one hidden layer, a so-called multi-layer perceptron (MLP). Compared to the
linear models, the MLP has an additional intermediate hidden layer $h$ with a
nonlinear activation function $f$. Thus, we first compute $\boldsymbol{h} = f
\left( \boldsymbol{W^1} \boldsymbol{x} + \boldsymbol{b^1} \right)$, and then
$\boldsymbol{y} = \boldsymbol{W^2} \boldsymbol{h} + \boldsymbol{b^2}$. The
output $\boldsymbol{y}$ is then scaled to the interval $(0,1)$ by the sigmoid
function $\sigma$ as in logistic regression and compared to the gold-standard
by cross-entropy. The gradient for updating the parameters is computed by the
chain-rule, also known as back-propagation. The optimization itself is carried
out by Adam~\cite{adam2014} with the default hyper-parameters and $\alpha=0.01$. We chose a hidden layer size of $1000$
and use rectified linear units~\cite{relu2010} as activation
function $f$ (except for the NYT dataset where we use $\tanh$ due to numerical
difficulties). For regularization, we apply dropout~\cite{dropout2012} with a
probability of $0.5$. The intermediate hidden layer can be regarded as a
fine-tuned task-specific word embedding, which enables the classifier as a
whole to learn nonlinear relationships among the features. To convert the odds
$\sigma (y)$ into a binary decision, several approaches suggest to use a
threshold learning technique~\cite{nam2014,DBLP:conf/www/TangRN09}. In our
initial experiments, however, we experienced that the most recent threshold
learning technique yields rather unsatisfactory results in terms of the $F_1$ measure. Instead, we use a fixed threshold of $0.2$. 
 
\subsubsection{Multi-Label Adaption}\label{sub:multi}

\paragraph{Binary Relevance} Linear models as well as Naive Bayes are
restricted to mutually exclusive class assignments by design. Only one class
out of all possible ones is selected. In multi-label classification, however,
multiple labels need to be assigned. The most common approach to adapt such
classifiers is to train one classifier per class, which distinguishes its
respective class from all others, \ie{} decides for binary
relevance~\cite{DBLP:journals/jdwm/TsoumakasK07} (also known as one-vs-all or
one-vs-rest). The training documents are supplied to all
label-specific classifiers. Depending on the prevalence of the label that
corresponds to the respective classifier, the example is either treated as
positive or as negative. At test time, the classification
result is composed of the binary decisions for each label.


\paragraph{Classifier Stacking}\label{sec:MVS}
Multi-value classification stacking~\cite{hess2008multi} refers to a technique where the final classification result is composed by two classifiers.
The so-called base-classifier returns a ranked list of label predictions with confidence scores.
Then for each class, a meta-classifier takes these confidence scores along with the position in the ranked list as input
and outputs a binary decision for the specific class. This technique enables
transforming any classifier that returns confidence scores into a multi-label classifier.
As meta-classifiers, we use decision trees with Gini impurity as splitting criterion.
To limit complexity, we generate training data only for those meta-classifiers,
whose class is among the top $30$ of the base-classifier's ranking~\cite{hess2008multi}.
We use this decision tree module (abbreviated with the suffix *DT) as an
alternative to hard cut-offs in Learning to Rank (see
\Cref{related_work}, and the fixed thresholds in multi-layer perceptrons
(see~\Cref{sub:eager}).  For comparison with the original work of He\ss{}
\etal{}~\cite{hess2008multi}, we also consider Rocchio as a base-classifier.
We furthermore experiment with applying the decision tree module on top of binary-relevance logistic regression.

\section{Experimental Setup}\label{sec:experiments}
We describe the datasets used for our experiments in \Cref{subsec:datasets},
before we outline the experimental procedure in \Cref{sub:procedure}.  We then
depict the conducted preprocessing and introduce our evaluation metric of a
sample-based $F_1$ measure in \Cref{sub:eval}.  We choose a sample-based
evaluation measure since it will assess the classification quality of each
document separately.  This reflects the workflow of manual document
classification as it is done by domain experts in scientific digital libraries
as well as journalists. 

\subsection{Datasets}\label{subsec:datasets}
  We have conducted our experiments on four datasets of English documents:
  two datasets are obtained from scientific digital libraries in the domains of economics and political sciences
  along with two news datasets from Reuters and New York Times. 
  Table~\ref{table:stats} summarizes the basic statistics of the datasets.
  For each document in the datasets, there are manually created gold-standard annotations provided by respective domain experts,
  who work as professional subject indexers in the corresponding organizations.
  %
  In addition, each dataset provides a domain-specific thesaurus that serves as controlled vocabulary of the gold-standard. 
  Its concepts are used as target labels in our multi-label document classification task.
  The thesaurus also offers sets of concept-specific phrases (\ie{}
  \texttt{skos:prefLabel} and \texttt{skos:altLabel} in case of SKOS
  format) that are used for concept extraction from the documents'
  full-text and titles~\cite{goossen2011news}. 
  %
  The \emph{economics} dataset consists of $62,924$ documents and is provided by ZBW --- Leibniz Information Centre for Economics.
  The annotations are taken from the Standard Thesaurus Wirtschaft (STW) version 9\footnote{\url{http://zbw.eu/stw/versions/9.0/about.en.html}}, which is a controlled domain-specific thesaurus for economics and business studies maintained by ZBW\@.
  The thesaurus contains $6,217$ concepts with $12,707$ concept-specific phrases.
  From these concepts, $4,682$ are used in the corpus and thus considered in the multi-label classification task. 
  Each document is annotated by domain experts with on average $5.26$ labels (SD\@: $1.84$).
  %
  The \emph{political sciences} dataset has $28,324$ documents.
  Similar to the economics dataset, we made a legal agreement for the political sciences dataset with the German Information Network for International Relations and Area Studies\footnote{\url{http://www.fiv-iblk.de/eindex.htm}} that is providing the documents.
  The labels are taken from the thesaurus for International Relations and Area Studies\footnote{\url{http://www.fiv-iblk.de/information/information_thesaurus.htm}}, which contains $9,255$ concepts (and an equivalent number of concept-specific phrases, i.\,e., there are no alternative phrases).
  From these concepts, $7,234$ are used in the corpus. 
  Each document in the dataset has on average $8.07$ labels (SD\@: $3.03$).
%
  The \emph{Reuters RCV1-v2} dataset contains $805,414$ articles.
  We chose articles where both the titles and the full-text of the documents are available.
  From this set of documents, we randomly selected $100,000$ articles to match
  the scale of the scientific corpora.
  In our experiments, we employ the thesaurus re-engineered from the Reuters dataset by Lewis \etal~\cite{lewis04}.
  The thesaurus contains $117$ concepts and a total of $173$ concept-specific phrases. 
  From these concepts, $101$ are used in the corpus. 
  Each document was annotated with on average $3.21$ (SD\@: $1.41$) labels.
%
  The \emph{New York Times Annotated Corpus Dataset} (NYT) contains $1,846,656$ articles.
  Each article has two sets of annotations,
  consisting of annotations created by a professional indexing service
  and annotations which were added by the authors using a semi-automatic system.
  We used the annotations provided by the indexing service because it is reasonable to
  expect that they are more consistent and of higher quality (cf.~\cite{hess2008multi}).
As for the Reuters dataset, we chose a random subset of $100,000$ documents containing both full-text and titles.
The number of concepts in the NYT dataset is $25,226$.
From these concepts, $6,809$ are used in our random sample.
Each document is annotated with on average $2.53$ (SD $1.78$) labels.
Like the political sciences dataset, each concept consists of only a single specific phrase.

\begin{table}
  \caption{Statistics for the datasets: $\left| D \right|$ documents, $\left|
    C \right|$ concepts in the thesaurus, $\left| L \right|$ labels assigned
    in the dataset, $d/l$ mean documents per label, $l/d$ mean labels per
    documents along with median ${l/d}_{50}$, $V$ vocabulary size, $w/d$ mean terms per document,
  and $c/d$ mean concepts per document}\label{table:stats}
  \resizebox{\columnwidth}{!}{%
    \begin{tabular}{lllll}
      \toprule
      & \textbf{Econ.} & \textbf{Polit.} & \textbf{RCV1} & \textbf{NYT} \\
      \midrule
      $\left| D \right|$ & $62,924$ & $27,576$ & $100,000$ & $100,000$\\
      $\left| C \right| $ & $6,217$ & $9,255$ & $117$ & $25,226$\\
      $\left| L \right| $ & $4,682$ & $7,234$ & $101$ & $6,809$\\
      $d/l$ & $70.8$ $(322.9)$ & $32.6$ $(116.8)$ & $3174.9$ $(6371.3)$ & $37.1$ $(213.0)$\\
      $l/d$ & $5.26$ $(1.84)$ & $8.57$ $(3.03)$ & $3.21$ $(1.41)$ & $2.53$ $(1.78)$\\
      ${l/d}_{50}$ & $4$ & $5$ & $14$ & $2$\\
      $V_\text{title}$ & $19,579$ & $15,419$ & $32,859$ & $40,736$ \\
      ${w/d}_\text{title}$ & $7.07$ $(3.03)$ & $8.13$ $(5.29)$ & $12.21$ $(2.39)$ &$4.46$ $(2.25)$\\
      ${c/d}_\text{title}$ & $3.33$ ($1.83$) & $3.69$ ($2.36$) & $0.57$ ($1.02$) & $0.70$ ($0.83$)\\
      $V_\text{full}$ & $1,340,628$ & $1,165,919$ & $155,339$ & $270,710$ \\
      ${w/d}_\text{full}$ & $6,750$ $(6,854)$ & $11,255$ $(15,565)$ &  $136$ $(114)$ & $310$ $(294)$ \\
      ${c/d}_\text{full}$ & $247$ ($121$) & $346$ ($189$) & $6.80$ ($8.60$) & $37.0$ ($38.2$)\\
      \bottomrule
  \end{tabular}}
\end{table}
  
\subsection{Procedure}\label{sub:procedure}

\paragraph{Vectorization methods}
We compare the different vectorization of the input text as shown in Figure~\ref{fig:pipeline} and described in \Cref{sub:vect}.
One vectorization is based on term frequencies (TF-IDF) and the other is based on concept frequencies (CF-IDF).
We experiment with the re-weighting method BM25 using term frequencies and BM25C using concept frequencies.
The concatenation of both terms and concepts is denoted by CTF-IDF and BM25CT, respectively.
As classifier, we employ kNN with cosine distance.
The performance of kNN relies on the assumption that documents are well represented by the features and that similar documents have similar labels.
Therefore, its classification performance is a good indicator for the quality of the features.

\paragraph{Classification methods} After determining the best-performing
vectorization method, we compare lazy learning as well as eager learning
classifiers of \Cref{sub:lazy,sub:eager} combined with the multi-label adaption
methods of \Cref{sub:multi}, where appropriate.  We leverage the linear models
(SVMs and logistic regression) to perform multi-label classification with
binary relevance, \ie{} training one classifier per label. To adapt the Learning to
Rank approach and the multi-layer perceptron to multi-labeling, we consider using
thresholds as well as stacking with decision trees.  We also experiment with stacking
the decision tree module on top of binary-relevance logistic regression.
Careful tuning of the hyperparameters is crucial to the success of machine
learning algorithms, especially in those multi-label classification tasks,
where only few training examples are available per class.  Striving to identify
well-suited hyperparameters that are invariant to the concrete dataset, we keep
all hyperparameters (as denoted in \Cref{sec:overview}) fixed across all experiments and datasets.

\subsection{Preprocessing and Evaluation}\label{sub:eval}

\paragraph{Preprocessing}
Prior to counting terms and extracting concepts, both the input text and the
concept-specific phrases of the thesauri are subject to preprocessing steps.
This includes discarding all characters except for sequences of alphabetic
characters with a length of at least two. Words connected with a hyphen are
joined (i.\,e., the hyphen is removed). Detected words were lower-cased and
lemmatized based on the morphological processing of WordNet~\cite{wordnet}. 

\paragraph{Evaluation} For evaluation, we separate each dataset into
90\% training documents and 10\% test documents and perform a $10$-fold
cross-validation, such that each document occurs exactly once in the test set.
Hence for each test document, we compare the predicted labels with the label
set of the gold standard and evaluate the $F_1$ measure. The $F_1$ measure is the
harmonic mean between precision, \ie{} true positives w.r.t false positives,
and recall, \ie{} true positives w.r.t false negatives. When no label is
predicted, the precision is set to zero. The F-scores are then averaged
over the test documents. We chose this sample-based $F_1$ measure over
class-averaged or global variants because it is closest to an assumed
application, where each individual document needs to be annotated as good as
possible. Please note, there is a possibility that all documents annotated
with a specific label fall only into one test set. Despite no training data is
available for these labels, we do \emph{not} exclude those
from our evaluation metric. Finally, we report the mean sample-based
F-score over the ten folds of the cross-validation. 
%

\section{Results}\label{sec:results}
In this section, we describe the results of our experiments. Due to the high
amount of possible pipeline configurations, we applied a step-by-step
approach. For both the text vectorization step and the classification step,
we search for a local optimum solution to find the best overall
classification strategy.

\paragraph{Results for Vectorization Methods}\label{sec:term_relevance}
Table~\ref{t-c-ct} shows the results for the text vectorization experiment.
The term-based vectorization method TF-IDF perform consistently better than the purely
concept-based vectorization CF-IDF methods on both the titles and the full-text.
The difference ranges from $0.003$ on Economics to $0.307$ F-score on Reuters.
When combining the term vector with the concept vector, the performance is at
least as good as the other text vectorization methods and in many cases yields
better results.
This is more noticeable on titles than on full-texts. 
BM25 re-weighting does not improve the results compared to TF-IDF neither in case of the titles nor the full-text. 
Rather, we observe a decrease in performance by up to 0.13.
These experiment using a nearest neighbor classifier indicates that CTF-IDF is the best-suited vectorization method.
Henceforth, we use CTF-IDF for comparing the performance of the classifiers.

\begin{table}[t]
\centering
\caption{Sample-averaged F-scores of the text vectorization methods with using kNN as common classifier}\label{t-c-ct}
\begin{tabular}{@{}llllll@{}}
\toprule
\textbf{Input}     & \textbf{Vectoriz.} & \textbf{Econ.} & \textbf{Polit.} & \textbf{RCV1}  & \textbf{NYT}\\
\midrule Full-text & TF-IDF            & 0.406          & 0.269           & 0.758          & 0.394\\
\midrule Full-text & BM25              & 0.370          & 0.230           & 0.740          & 0.370\\
\midrule Full-text & CF-IDF            & 0.402          & 0.266           & 0.451          & 0.367\\
\midrule Full-text & BM25C             & 0.296          & 0.161           & 0.423          & 0.236\\
\midrule Full-text & CTF-IDF           & \textbf{0.411} & \textbf{0.272}  & \textbf{0.761} & \textbf{0.406}\\
\midrule Full-text & BM25CT            & 0.377          & 0.231           & 0.742          & 0.379\\
\midrule
\midrule Titles & TF-IDF  & 0.351          & 0.201          & 0.709          & 0.238\\
\midrule Titles & BM25    & 0.349          & 0.196          & 0.687          & 0.230\\
\midrule Titles & CF-IDF  & 0.303          & 0.183          & 0.275          & 0.105\\
\midrule Titles & BM25C   & 0.304          & 0.172          & 0.193          & 0.073\\
\midrule Titles & CTF-IDF & \textbf{0.368} & \textbf{0.212} & \textbf{0.717} & \textbf{0.242}\\
\midrule Titles & BM25CT  & 0.364          & 0.208          & 0.693          & 0.239\\
\bottomrule
\end{tabular}
\end{table}
       
\paragraph{Results for Classifiers}

 
The results of comparing the different classifiers are documented in
\Cref{classifer_table}. 
As shown in the table, Bernoulli Bayes has a slight advantage over multinomial
Bayes for titles. On the other hand, the multinomial variant has a slight disadvantage
on full-texts. However, both methods consistently fall far behind kNN on
full-texts. In the case of working with titles, the Bayes classifiers are able
to keep up with kNN on two datasets. RocchioDT's scores are depending on the
datasets and range from the lowest (Reuters) to a score only slightly different
from kNN (NYT, political sciences). The generalized linear models SVM and
logistic regression are close to each other. The difference is no more than
$0.04$ for any dataset. Considering Learning to Rank, we observe that the technique yields 
consistently lower scores than the multi-layer perceptron.
Overall, the eager learners SVM, LR, L2R and MLP outperform both
Naive Bayes and the lazy learners Rocchio, and kNN\@. Among all classifiers,
MLP dominates on all datasets apart from NYT on titles, where LRDT achieves 
a $.021$ higher score. 
While the stacked decision tree module increases the F-scores of logistic
regression on all datasets with fewer than $100$ documents per label (all but
Reuters), the impact of the stacking method is inconsistent for the Learning to
Rank and MLP approaches.
It is noteworthy that there are cases where a classifier performs better on the title data
than the same classifier applied on the full-text data.  These are Bernoulli
Bayes on the Reuters dataset and RocchioDT on the economics dataset.  As a
general rule, however, full-texts generate higher scores than the titles.
Comparing different classifiers across titles and full-text, we can make the
observation that some classifiers trained on titles outperform others that were
trained on the full-text. Apart from the NYT corpus, the eager learners LR,
LRDT and MLP on titles are superior to kNN on full-texts.
Finally, we compare the F-scores of the best-performing multi-layer perceptron
on titles with its scores obtained on full-text. On the NYT dataset,
58\% of the F-score is retained when using only titles.
On the political sciences and economics datasets, the retained F-score is 83\% and 91\%, respectively.
On the Reuters dataset, the MLP using solely titles retains 95\% of the F-score
that is obtained with full-text information available.

\begin{table}[!h]
\centering
\caption{Sample-averaged F-scores for classification methods with using the best vectorization method CTF-IDF}\label{classifer_table}
\resizebox{\columnwidth}{!}{%
\begin{tabular}{@{}lllllll@{}}
\toprule
\textbf{Input}      & \textbf{Classifier} & \textbf{Econ.} & \textbf{Polit.} & \textbf{RCV1} & \textbf{NYT} \\
\midrule Full-text & kNN (\emph{baseline})      & 0.411              & 0.272                  & 0.761                & 0.406    \\
\midrule Full-text & Bayes (Bernoulli)   & 0.318              & 0.191                  & 0.657                & 0.281    \\
\midrule Full-text & Bayes (Multinom.) & 0.235              & 0.207                  & 0.703                & 0.349    \\
\midrule Full-text & SVM                 & 0.481              & 0.319                  & 0.852                & 0.554 \\
\midrule Full-text & LR                  & 0.485              & 0.322                  & 0.851                & 0.556    \\
\midrule Full-text & L2R                 & 0.431              & 0.328                  & 0.727                & 0.435    \\
\midrule Full-text & MLP                 & \textbf{0.519}     & \textbf{0.373}         & \textbf{0.857}       & 0.569 \\
\midrule Full-text & RocchioDT           & 0.291              & 0.225                  & 0.645                & 0.393    \\
\midrule Full-text & LRDT                & 0.498              & 0.339                  & 0.843                & 0.562    \\
\midrule Full-text & L2RDT               & 0.415              & 0.280                  & 0.751                & 0.421    \\
\midrule Full-text & MLPDT               & 0.492              & 0.340                  & \textbf{0.857}       & \textbf{0.578} \\
\midrule
\midrule Titles & kNN                 & 0.368          & 0.212          & 0.717          & 0.242    \\
\midrule Titles & Bayes (Bernoulli)   & 0.301          & 0.179          & 0.708          & 0.233    \\
\midrule Titles & Bayes (Multinom.) & 0.254          & 0.178          & 0.699          & 0.214    \\
\midrule Titles & SVM                 & 0.426          & 0.272          & 0.804          & 0.325 \\
\midrule Titles & LR                  & 0.429          & 0.274          & 0.803          & 0.326    \\
\midrule Titles & L2R                 & 0.419          & 0.296          & 0.699          & 0.296    \\
\midrule Titles & MLP                 & \textbf{0.472} & \textbf{0.309} & \textbf{0.812} & 0.332   \\
\midrule Titles & RocchioDT           & 0.335          & 0.219          & 0.584          & 0.252    \\
\midrule Titles & LRDT                & 0.451          & 0.279          & 0.796          & \textbf{0.353} \\
\midrule Titles & L2RDT               & 0.428          & 0.261          & 0.730          & 0.25  \\
\midrule Titles & MLPDT               & 0.457          & 0.277          & 0.808          & 0.340    \\
\bottomrule
\end{tabular}}
\end{table}


\section{Discussion}\label{sec:discussion}
 
The results show that multi-label classification of text documents can be
reasonably conducted using only the titles of the documents. Over all datasets,
the multi-layer perceptron on titles retains 82\% of the F-score obtained on
full-text. This gives an empirical justification for the value of automated
semantic document annotation using metadata. 
From the first experiment, we find that combining words with extracted concepts
as features is preferable over one of them alone.
Concepts hold valuable domain-specific semantic information. The term
frequency on the other hand, holds implicit information which is as well important
for correct classification.
Eager learners are, by design, capable of learning which terms or concepts need
to be associated to the respective class. The results show that also lazy
learners benefit from this joint representation.
The second experiment shows that eager learners such as logistic regression and
MLP consistently outperform lazy learners for multi-label classification.  This
result extends recent advancements in multi-labeling~\cite{nam2014,TANAKA201585} towards document classification
scenarios with many possible output labels and only few examples per class.

Inspecting the results for titles and full-text, the best-performing
classifiers still perform better on the full-text. This is not surprising
since the full-text holds considerably more information (including the title). However,
for all datasets apart from the NYT dataset, the difference in F-score of the
best-performing MLP is small. 
The difficulties in classifying the documents in the NYT dataset can be explained by a characteristic that the titles consist only of $4$ words on average. 
There may be a lower bound on the title length to perform the classification task, since a short title limits the amount of available information and thus prohibits discrimination. 
From the other datasets, we can state that an average of $7$ words per title leads to at least 80\% retained F-score.
Thus, it would require further investigation to understand the specific influence of the title length on the classification performance.
The complexity of a multi-labeling problem depends on the number of available
documents per label, independent of whether the full-text or the titles are
used. Especially binary-relevance classifiers suffer from conservative label
assignments (high precision, low recall), when many negative examples and only
few positive examples are presented during training.
While the results of the stacked decision tree module are inconsistent for MLP and L2R,
it does alleviate the conservative assignments problem of binary-relevance,
when only few documents per label are available.

In our experiments over four large-scale real-world corpora covering a broad
range of domains (economics, political sciences and  news), we did not limit
the complexity by excluding rare labels and kept all independent variables as
well as hyperparameters fixed.
In our prior work~\cite{grosse2015comparison}, we have used the
thesaurus hierarchy to model label dependencies which improves the classifications obtained by kNN.
Despite not making use of the hierarchy anymore, we are able to achieve even
higher absolute F-scores using eager learning techniques and supplying
term features in addition to extracted concepts. We can therefore drop the
constraint of a hierarchical organization among the labels.
Due to this minimal amount of requirements and invariant configurations of the text processing pipeline, we
can expect our findings to generalize to a wide range of other corpora.

To validate the practical impact of the experimental results, we have conducted
a qualitative assessment of the experimental results in an expert workshop with
three subject indexing specialists at ZBW, the national library for economics in
Germany.  The experts state that titles can be sufficient for classification of
scientific documents.  They further noted that titles contain less
information than what an intellectual indexer has available when manually
conducting the classification tasks for the documents. They also pointed out
that researchers carefully chose their titles for findability. The experts
argued that reasonably good automatic indexing based on titles is valuable
since it does not raise legal problems compared to processing full-text as
discussed in the introduction. We conclude that using the documents' title for
automated semantic annotation is not only technically possible with a high
quality but also valuable from a practical point of view.

\section{Conclusion}

We have shown that it is reasonable to conduct semantic annotations of documents by just analyzing the titles.
Our experiments show that by using titles, a performance of over 90\% can be reached w.r.t to the classification performance obtained when using the full-text of the documents.
This opens many new possibilities for using document classification even when only little input data is available such as titles obtained from the documents' metadata on the Linked Open Data cloud.
 
To encourage further research in the field and to invite other
researchers to compare and develop further methods, the full source code of our
generic text processing pipeline is available on GitHub\footnote{\url{https://github.com/Quadflor/quadflor}}. 
We invite practitioners and developers to use and extend the framework.


\paragraph*{Acknowledgement}
This research was co-financed by the EU H2020 project MOVING (\url{http://www.moving-project.eu/}) under contract no 693092.
We thank Tobias Rebholz, Gabi Schädle, and Andreas Oskar Kempf from ZBW for valuable discussions in our expert workshops on how to use SKOS vocabularies and titles to annotate scientific papers.

\bibliographystyle{ACM-Reference-Format}
\bibliography{quadflor-brief.bib}
\end{document}